\documentclass[preprint,superscriptaddress,10pt]{revtex4-2}
\usepackage{xspace} 
\usepackage{makecell} %\thead
\usepackage{siunitx}
\usepackage[version=4]{mhchem}
\usepackage{color,soul}
\usepackage{graphics}
\usepackage{bm}        % for math
\usepackage{amssymb}   % for math
\usepackage{amsfonts}
\usepackage{amsmath,mathtools} % mathtools to use \begin{multlined}
\usepackage{multirow}
\usepackage{array}
\usepackage{booktabs}
\usepackage[export]{adjustbox}
\usepackage{textcomp}%for \textdegree
\usepackage{gensymb}
\usepackage{lipsum}
\usepackage{graphicx}
\usepackage{setspace}
\usepackage{epstopdf}
\DeclareGraphicsExtensions{.eps}
\newcommand{\mos}{MoS$_2$\xspace}
\begin{document}
	\setstretch{1.2}
	\title{Experimental demonstration of two distinct pathways of trion generation in monolayer MoS$_2$}

	\author{Faiha\ Mujeeb}
	\affiliation{Department of Physics$,$ Indian Institute of Technology Bombay$,$ Mumbai 400076$,$ India}
	\author{Arkaprava\ Chowdhury}
	\affiliation{Department of Chemistry$,$ Indian Institute of Technology Bombay$,$ Mumbai 400076$,$ India}
	\author{Anindya\ Datta}
	\affiliation{Department of Chemistry$,$ Indian Institute of Technology Bombay$,$ Mumbai 400076$,$ India}
	\author{Subhabrata\ Dhar}
	\email{dhar@phy.iitb.ac.in}
	\affiliation{Department of Physics$,$ Indian Institute of Technology Bombay$,$ Mumbai 400076$,$ India}

	\begin{abstract}
 Excitation power and energy dependent  photoluminescence (PL) and transient absorption spectroscopy (TAS) studies are carried out on  chemical vapour deposition (CVD) grown 1L-\mos films to understand the process of trion formation.  The study shows  that the excitation with sufficiently low photon energy results in the creation of trions  directly in the K/K$^\prime$ valleys through photon absorption followed by  phonon scattering events. On the other hand, excitation energy sufficiently larger than the band-gap can generate the carriers away from the K/K$^\prime$ valleys.  Dissimilarity in the rates of relaxation of the photo-excited electrons and the holes to the bottom of the K/K$^\prime$ valleys results in the transformation of the excitons residing there into trions. Our TAS study clearly demonstrates a temporary increase of the trion population in the K/K$^\prime$ valleys.  Moreover, excitation intensity dependent PL spectroscopy performed under above-band-gap excitation, also suggests the  coexistence of both the pathways of trion generation in this material. This conclusion is further validated by a rate equation model.  Our findings provide valuable insight into the formation of trions in monolayer transition metal dichalcogenides (TMDC), which could be crucial in designing  valleytronic devices based on trions .
	\end{abstract}
	
	\maketitle  
	
	\section{Introduction}

Monolayer (1L) transition metal dichalcogenides (TMDCs) have garnered significant interest due to their unique optical and electronic properties \cite{mak2010atomically,mak2013tightly,mak2012control,Abhay2023novel,Swarupdeb2022cumulative}. As compared to the bulk semiconductors and semiconductor heterojunction based quasi-2D systems, the 2D confinement of carriers leads to an order of magnitude higher binding energy of excitons in these materials \cite{qiu2013optical,mak2013tightly,kheng1993observation,huard2000bound}, which  results in various  interesting excitonic phenomena, such as the formation of excitonic complexes like  trions (charged excitons), biexcitons (complex of two excitons) and charged biexcitons (complex of a trion and an exciton),  at relatively much higher temperatures \cite{plechinger2015identification,berghauser2014analytical,faiha2024solving}.  Trions with positive and negative polarities can be formed  in the presence of excess electrons and holes, respectively.  Note that due to much lower binding energies, trions can only be found at extremely low temperatures in bulk semiconductors and quasi-2D systems \cite{kheng1993observation,huard2000bound}. Whereas, with the binding energy in the range of 10-43 meV \cite{plechinger2015identification,mak2013tightly,jones2013optical,ross2013electrical,singh2014coherent,zhang2015valence,christopher2017long}, trions in 1L-TMDC  can be observed  even at room temperature \cite{jones2013optical,ross2013electrical}. This makes these monolayers an excellent platform for studying the properties of trions. These charged excitonic entities are especially interesting for device applications because their binding energy and the population can be tuned either via chemical doping or gate biasing \cite{bellus2015tightly,luo2020gate,das2020gate,mouri2013tunable,Faiha_AIP}. Moreover,  utilising their charge character a trion current can also be electrically set-up and controlled  in the medium. Like excitons, trions, in TMDC monolayers, also exhibit valley polarization (VP) effect \cite{Faiha2023_PRB},  which enables an exciton/trion to sustain its valley character throughout the time of its existence. It is also important to note that the process of recombination of trion is not as favourable as that of an exciton as the  recombination of the electron-hole pair  in a negative (positive) trion leaves the extra electron (hole) at a state that is several hundreds of meV (excitonic binding energy) below the band edge \cite{strain_polarization_zhu,TrionPol_singh,Faiha2023_PRB}. This scenario is not energetically beneficial  as  the left-out electron (or hole) needs to gain additional energy to reach an allowed state in the band structure. Trions can thus survive longer than the excitons, which has indeed been observed experimentally  in 1L-TMDC \cite{chang_trionTRPL,gao2016valley_trionhighpower,wang2013valley_ACS,yan2017long}.   All these properties make trion a potential candidate for future excitonic and valleytronic devices, which are expected to be faster and better compatible with the optical communication than the conventional electronics.  Understanding trions in 1L-TMDCs has thus gained a lot of impetus \cite{golovynskyi2021trion,mak2012control,lui2014trion}.  However, there is still a lot to be comprehended. In particular, the mechanism of trion generation and recombination is yet to be fully understood.

Here, we have carried out excitation power and energy dependent photoluminescence (PL) and transient absorption spectroscopy (TAS)  in chemical vapour deposition (CVD) grown 1L-\mos samples to understand the trion generation mechanism. Our study reveals that trions can be generated at the bottom of the K/K$^\prime$ valleys in two pathways. In the pathway-1, trions are formed through photon absorption, which may be followed by phonon scattering. Whereas in the pathway-2, carriers are photo-generated to the higher lying valleys of the conduction and valence bands and later relax to the K/K$^\prime$ valleys to form trions. It has been found that  under the excitation with photon energy closely matching the transition energy of the $A$-excitons, trions are primarily formed via the pathway-1.  Whereas, the coexistence of the two trion generation processes can be seen when the excitation is carried out with the above band-gap light. To further substantiate these observations, a rate equation model involving the generation and recombination of excitons, trions, and free excess carriers, as well as exciton-trion inter-conversions is set-up, which validates our conclusions.

 \section{Experimental Techniques}
 Large area covered 1L-\mos films were grown on sapphire substrates by CVD technique \cite{growth}. The monolayer films were later transferred to SiO$_{2}$/Si wafers by polystyrene assisted transfer procedure \cite{transfer_gurarslan}. The transferred samples were found to have a higher population of trions as compared to the as-grown films \cite{chakrabarti2022enhancement,Faiha_AIP}. This was attributed to n-type doping resulting from the leftover polystyrene capping \cite{Faiha_AIP}. Power dependent PL spectra were recorded at 80~K under three different excitation energies (2.33 eV and 2.04 eV DPSS laser, 1.96 eV HeNe laser). In all measurements, Linearly polarized lasers are used. The data were collected in a microscopy set-up equipped with a 50$\times$ objective lens in a back scattering geometry. The spectra were recorded using a 0.55~m focal length monochromator equipped with a Peltier cooled CCD detector. Transient absorption spectroscopy (TAS) measurements were performed using a femtosecond Ti-Sapphire laser of wavelength of 800 nm, pulse duration of 100 fs, and repetition rate of 1 kHz.  The 600~nm (2.06~eV) and 400~nm (3.1~eV) pump pulses were used for the study. The 600~nm pump was generated using an Optical Parametric Amplifier (TOPAS, Light Conversion, Lithuania), while  400~nm pump is generated via frequency multiplication of 800~nm laser.  Part of the 800 nm light was passed through a CaF$_{2}$ window to generate a white light continuum (370–750 nm). This white light was split into probe and reference beams using a beam splitter. The reference beam was used to correct for intensity fluctuations of the probe beam. These measurements were conducted at 295~K and 80~K inside a liquid nitrogen cryostat. 
  
 \section{Results and Discussion}
\begin{figure*}[htb]
	\includegraphics[scale=1]{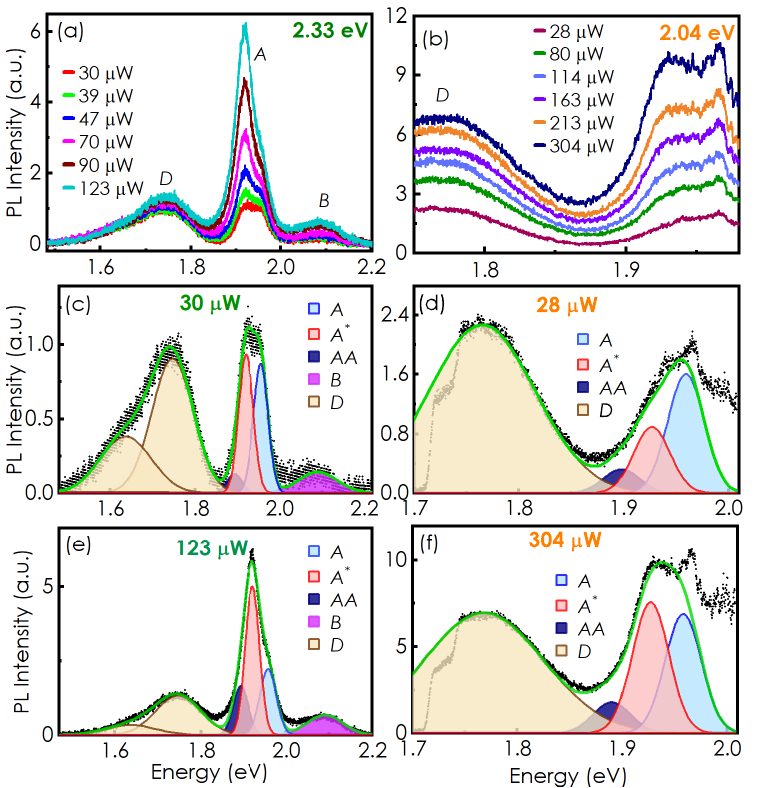}
	\caption{PL spectra recorded at 80~K under different excitation powers for excitation photon energy of (a) 2.33~eV (532~nm) and (b) 2.04~eV (607~nm). Gaussian deconvolution of the spectra for excitation photon energy of  2.33~eV with excitation power of (c) 30~$\mu$W and (e) 123~$\mu$W as well as for excitation photon energy of  2.04~eV  with excitation power of (d) 28~$\mu$W and (f) 304~$\mu$W. The sharp features at $\sim$1.965~eV are the Raman peaks resulting due to resonant excitation. The photons below 1.75~eV and above 1.98~eV is cut off by filter used to remove laser line. The black symbols represent the experimental data and the green lines stand for the cumulative fit.}
	\label{fig:trionformpl}
\end{figure*}

\begin{figure}[htb]
	\includegraphics[scale=0.95]{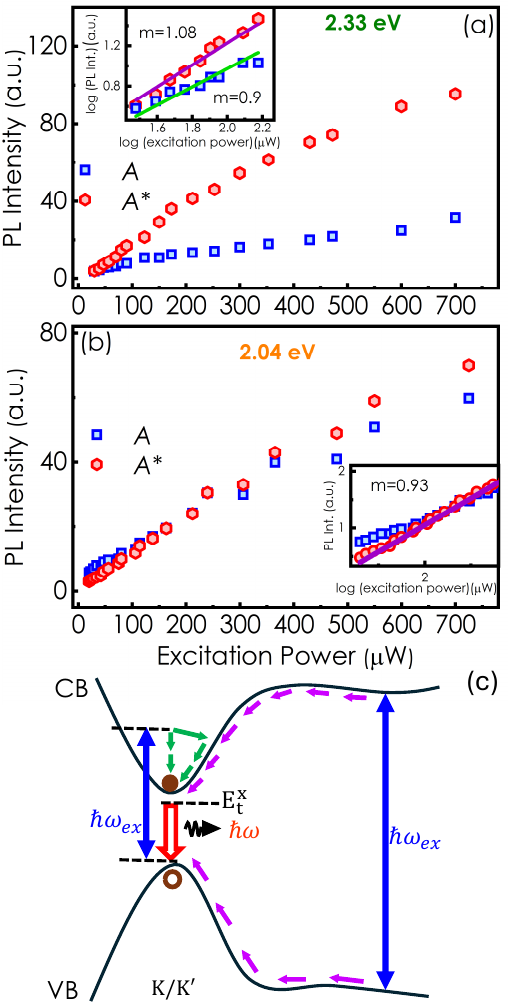}
	\caption{Integrated PL intensity as a function of excitation power for excitons (A, blue symbols) and trions (A$^\ast$, red symbols)  for excitation photon energy of (a) 2.33~eV and (b) 2.04~eV. Insets show the log-log plots.(c) Schematic depiction of generation of trions. }
	\label{fig:ratio}
\end{figure}

The PL spectra  recorded at 80~K for different excitation powers and at photon energies 2.33 and 2.04~eV are presented in Fig.~\ref{fig:trionformpl} (a) and (b), respectively.  Spectra collected at even higher excitation powers (up to 700 $\mu$W) can be found in the supplementary material Fig.~S1 \cite{sup}. In case of Fig.~\ref{fig:trionformpl} (a), the $A$-excitonic feature appearing at $\sim$1.92~eV clearly shows two distinct peaks, which correspond to the neutral excitons and the trions. This becomes clearer when the spectra are  deconvoluted with six gaussian functions representing neutral exciton ($A$), trion ($A^\ast$), biexciton ($AA$), $B$-exciton ($B$), and two defect peaks ($D$) \cite{tailoring,faiha_2024recombination}. Deconvolution for the spectra recorded under above band-gap excitation photon energy of 2.33~eV for two excitation powers are shown in Fig.~\ref{fig:trionformpl} (c) and (e). It is evident that at the lowest excitation power of 30~$\mu$W, the peak intensities of excitons and trions  are almost the same.  While at 123~$\mu$W, the trion peak clearly dominates over the excitonic one. It has been found that the trion emission intensity is increasing at a much faster rate with the excitation power than that of the excitons.  At a photon energy of 2.33~eV, the carriers can be excited deep inside the conduction and valence bands in this material. 

Under the excitation energy of 2.04~eV, which is less than the band-gap energy but slightly more than the  transition energy of $A$-excitons in 1L-MoS$_2$,  the variation of the PL spectra with the excitation power exhibits significantly different trend as shown in  Fig.~\ref{fig:trionformpl} (b). Note that the sharp features at $\sim$1.965~eV are the Raman peaks resulting due to resonant excitation.  Evidently, the excitonic feature  shows a clear red shift as the excitation power increases. Note that the laser cut-off filter has a band-pass window from 1.75 to 1.98~eV, which removes both the high energy tail of the $A$-excitonic feature and the low energy part of the defect feature.  These spectra are deconvoluted with four gaussians representing the neutral exciton ($A$), trion ($A^\ast$), biexciton ($AA$), and only one defect  feature ($D$).  It should be mentioned that Raman features appearing at the high-energy edge of the spectra are avoided in the fitting process. The deconvolution clearly demonstrates that as the excitation power increases, the trion-to-exciton peak intensity ratio enhances resulting in a red-shift of the peak [see Fig.~\ref{fig:trionformpl} (d) and (f)].  The study carried out with the excitation photon energy of 1.96~eV, which is resonant with $A$-exciton transition energy  in 1L-MoS$_2$, shows the similar trend as obtained with the excitation energy of 2.04~eV [see supplementary material Fig.~S2 \cite{sup}]. 

%The results are shown for the spectra recorded at 28 and 304~$\mu$W of excitation powers are shown in Fig.~\ref{fig:trionformpl} (d) and (f), respectively, as examples.  deconvolution of the spectra  (). To further validate this observation, measurements were performed under under 1.96~eV (633~nm), which resonant to A exciton at 80~K (Fig.~S2, supplementary material). At 80 K, the peak maximum could not be identified clearly due to resonance effects, so spectra were recorded at 100~K. Increasing the temperature further could reduce the trion population \cite{tailoring}. The two sharp features at $\sim$1.91~eV are Raman peaks. Inset shows the A exciton complex at lowest powers. The spectra behave very similarly to that of 2.04~eV excitation. Both 2.04~eV and 1.96~eV are sub bandgap excitation.

        The integrated PL intensities both for trions and excitons are plotted as functions of the excitation power in Fig.~\ref{fig:ratio} (a) and (b) for the excitation  photon energy of 2.33 (above band-gap) and 2.04~eV (below band-gap), respectively. It is interesting to note that under 2.33~eV excitation, trion contribution is significantly higher than that of the excitons.  Moreover, both the species exhibit a faster followed by a slower rate of intensity increase with the excitation power.  The PL intensity variation with the excitation power is linear in the two regimes both for the excitons and the trions, which  is further demonstrated in the log-log plot of the data shown (only up to 160 $\mu$W of excitation power) as the inset of the Fig.~\ref{fig:ratio} (a). The data can be fitted with straight lines with slope $m$ = 1.08 for the trions and $m$ = 0.9 for the excitons. Both the values of  m are very close to 1.  In contrast, under 2.04 eV excitation, the intensity versus excitation power data can be fitted with a single line both for trions and excitons. The two plots also closely mirror each other.  The linearity is further confirmed in the log-log plot of the data shown as the inset of the Fig.~\ref{fig:ratio} (b). The data can be fitted with a straight line of slope $m$ = 0.93, which is very close to 1.  The apparent deviation of the excitation power dependence of the exciton intensity from linearity at low excitation powers might be due to the overestimation of the excitonic contribution as the relative strength of the resonant Raman peak (also appearing at the energy location) becomes significant at low powers [see Fig.~\ref{fig:trionformpl} (b), (d) and (f)] At any given excitation power, the trion-to-exciton intensity ratio is more for the excitation of 2.33~eV  than  2.04~eV. It should be noted that enhancement of this ratio has earlier been reported in 1L-\mos, when the excitation energy is changed from 2.33 to 3.06~eV (both the photon energies are more than the band-gap of 1L-\mos) \cite{golovynskyi2021trion}. These observations can be attributed to the two distinct pathways for trion formation. Since the momentum of photons is negligible, conservation of energy and momentum demands the excitons/trions to be at the valley extrema (bottom of the valley) for the radiative recombination to take place. In the pathway-1, both trions and excitons can be formed at the centre of the K/K$^\prime$ valleys via absorption of photons followed by phonon scattering as schematically shown in Fig.~\ref{fig:ratio} (c) with the green arrows. Note that when the excitation energy is resonant with the transition energy of the trions (excitons), involvement of phonons is not necessary. Formation of trions additionally requires the presence of excess carriers in these valleys.  In the pathway-2, free carriers can be generated away from the K/K$^\prime$ valleys  when excited with above-band-gap light of sufficiently high energy. These photo-generated highly energetic electrons and holes move towards the K/K$^\prime$ valleys as shown by the purple arrows in Fig.~\ref{fig:ratio} (c), which can lead to the formation of excitions at the bottom of these valleys. The difference in the rate of arrival of the two carrier types can create excess population of either electrons or holes in the K/K$^\prime$ valleys, which in turn facilitates the conversion of excitons into trions.  Note that the  transformation of excitons into  trions  with a time-scale of  a few picoseconds has been experimentally observed in exfoliated 1L-MoSe$_2$ \cite{trion_dynamics}. Since under 2.33 eV excitation, both pathways of trion generations are possible, we see two different regimes of intensity variation with the excitation power [Fig.~\ref{fig:ratio} (a)].  The initial increase  at a faster rate followed by the slower enhancement  can be attributed to the pathway-1 and pathway-2 dominated regimes, respectively. We believe that at sufficiently high excitation powers, the rate of increase of the steady state population of both excitons and trions with excitation power suppresses in ‘pathway-1’ due to higher rate of nonradiative recombination of these entities via exciton-exciton annihilation effects. As pathway-1 suppresses at higher values of the excitation power, ‘pathway-2’ gradually becomes the dominant trion generation mechanism. On the other hand, under 2.04~eV excitation, excitons/trions are formed only through pathway-1. This is why we observe only a single rate of variation of the exciton/trion intensity over the entire range of excitation power.

    Excitons can be formed in both geminate as well as bimolecular pathways within the K/K$^\prime$ valleys when illuminated with above band-gap photons \cite{piermarocchi1997exciton}. In geminate route, an exciton can be generated at some point on its energy versus center-of-mass-momentum ($\hbar K$) dispersion profile through simultaneous absorption of a photon and emission/absorption of phonons with wavevectors satisfying the momentum conservation. On the other hand, in bimolecular formation pathway, photoexcited hot carriers (electrons and holes) are first thermalized through carrier-phonon and carrier-carrier scattering processes and then they interact to form excitons. When the excitation photon energy is more than the excitonic transition energy but less than the bandgap, geminate formation is the only pathway for generating excitons. However, under above band-gap excitation, both the pathways can be active. The rate of geminate formation process should be proportional to $P$, the density of photo-generated electron-hole pairs ($\propto$ excitation intensity $G$). While the bimolecular formation rate of excitons should be proportional to the product of thermalized electrons and holes, which will again be proportional to $G$, if thermalization happens at a much faster rate than the recombination rates of the excitons. Note that the formation time of excitons in TMDC monolayers are experimentally estimated to be as small as sub-100 fs \cite{ceballos2016exciton,steinleitner2017direct,trovatello2020ultrafast}. It has also been demonstrated that the rate of neutral exciton generation in these monolayers  increases linearly with $G$ \cite{steinleitner2017direct,trovatello2020ultrafast}.  

 When illuminated with photons, whose energy is such that the photogeneration is restricted within the K/K$^\prime$ valleys, trions can be formed within the K/K$^\prime$ valleys through pathway-1 in geminate as well as two-particle and three-particle formation routes . In the geminate route, a trion can be generated from the photo-generated electron-hole pair and an extra carrier (electron or hole) at some point on the energy-$K$ dispersion curve of the trion through simultaneous absorption of a photon and emission/absorption of phonons satisfying the energy and momentum conservation. The rate of this process will be proportional to the product of $P$ (and hence $G$) and $n$ (excess free electron (hole) concentration in K/K$^\prime$ valleys). These excess carriers can result from the unintentional back-ground doping (for S-vacancy can act as shallow donors in \mos) and the dissociation of trions into excitons and unbound carriers. In two-particle process of trion formation, excitons are first formed and then they interact with excess electrons to form trions. While in three-particle formation process, trions can be generated from the thermalized unbound carriers.  Due to very strong carrier-carrier and carrier-LO phonon scattering processes in TMDC monolayers, formation time of both excitons and trions are expected to be extremely small. In fact, the formation time of excitons and trions in TMDC monolayers are experimentally estimated to be as small as sub-100 fs (as mentioned before) and sub-ps \cite{gao2016valley_trionhighpower,nakama2024trion}, respectively. Both the rates are order of magnitude faster than the decay rate of exciton and trion densities (time scale of more than several tens of ps). Since the formation of excitons as well as the thermalization of excess carriers are expected to be extremely fast process, the bimolecular formation of trions in the K/K$^\prime$ valleys can also be considered to be proportional to $nG$. When the excitation energy is sufficiently high such that photo-carriers can also be generated in higher lying valleys, which leads to the formation of trions in the K/K$^\prime$ valleys through pathway-2 as discussed before. The rate of this process should be proportional to the product of $n$ and the exciton concentration $N_{x}$.

The populations of excitons ($N_{x}$), trion  ($N_{t}$), and excess free carriers ($n$) can thus be expressed by the following set of rate equations,  

\begin{align}
	\frac{dN_{x}}{dt} &=\alpha_{x}G -\gamma_{x}N_{x}+\beta_{tx}N_{t}-\beta_{xt}^{\prime}nN_{x}  \nonumber  \\
	\frac{dN_{t}}{dt} &=\alpha_{t}nG -\gamma_{t}N_{t}-\beta_{tx}N_{t}+\beta_{xt}^{\prime}nN_{x} \label{eqnn}   \\
	\frac{dn}{dt} &=\alpha_{n}G +\beta_{tx}N_{t}-\beta_{xt}^{\prime}nN_{x}-\alpha_{t}nG   \nonumber 
\end{align}

Where,  $\alpha_{x}$,  $\alpha_{t}$ and  $\alpha_{n}$ are the proportionality constants associated with the generation of excitons, trions and excess carriers (electrons or holes), respectively. $\gamma_{x}$ and $\gamma_{t}$  are the recombination rate-constants for the excitons and trions, respectively. Note that each one represents the sum of the rate-constants associated with both the radiative and nonradiative recombination processes. $\beta_{tx}$ and $\beta_{xt}^{\prime}$ are the coefficients for the rates of trion-to-exciton and exciton-to-trion conversions, respectively.  In trion-to-exciton conversion process, one trion is dissociated to form an exciton and an excess carrier. It should be noted that $n(t)$ = $n_o$ + $n_{ph}(t)$, where $n_o$ represents the background electron concentration arising from the background doping and $n_{ph}(t)$ accounts for the photo-induced change in electron concentration. Under the steady state condition, $\frac{dN_{x}}{dt}=\frac{dN_{t}}{dt}=\frac{dn}{dt}=0$ and the above equations results in a rather simple  solution for the excess carrier concentration $n$. 
\begin{equation}
	n = \frac{\alpha_{n}G+\beta_{tx}N_{t}}{\beta_{xt}^{\prime}N_{x}+\alpha_{t}G}
\end{equation}
 The set of equations can be solved in two different regimes: (1), when trion formation of  rate in pathway-1 dominates over pathway-2 ($\alpha_{t}G > \beta_{xt}^{\prime}N_{x}$) and (2), the pathway-2 dominated regime, when $\beta_{xt}^{\prime}N_{x} > \alpha_{t}G$. In both the regimes, it can be shown that trion and exciton populations are proportional to the generation rate (see supplementary material S3). However, the proportionality constants are different in the two regimes, which are indeed consistent with the observations of  Fig.~\ref{fig:ratio} (a) and (b). 
 
 \begin{figure}[htb]
	\includegraphics[scale=0.92]{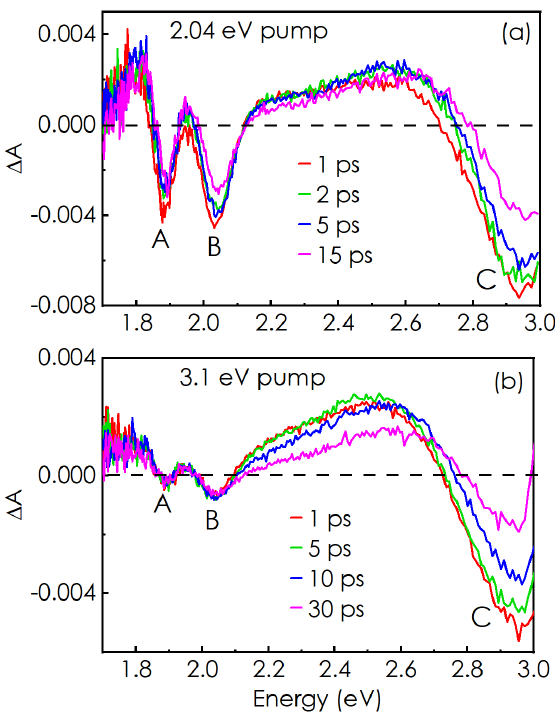}
	\caption{Transient absorption spectra recorded after different delay times under pump energy of (a) 3.1 eV and (b) 2.06 eV with 1 mW power at room temperature. The black dashed line represent $\Delta$A $=$ 0.}
	\label{fig:TAS_abs}
\end{figure} 

\begin{figure}[htb]
	\includegraphics[scale=1]{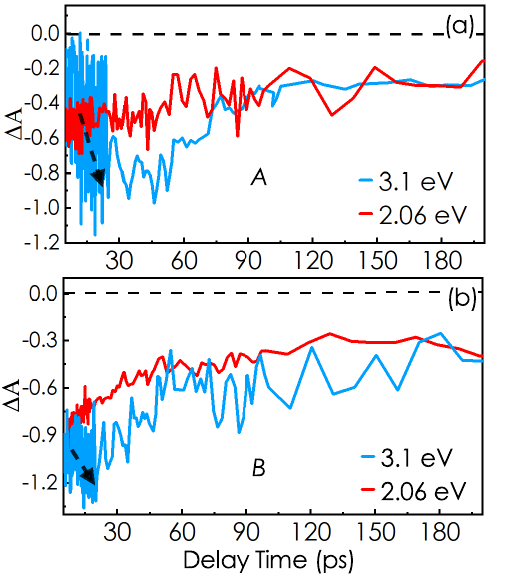}
	\caption{Normalized $\Delta$A versus delay time profiles for (a) $A$-excitons and (b) $B$-excitons recorded at room temperature with 3.1~eV (Blue) and 2.06~eV (Red) pump.  Dashed (black) arrows in the figures  mark the increase of $\Delta$A.}
	\label{fig:trionformtas1}
\end{figure}

To further elucidate the mechanisms behind trion generation, ultrafast pump-and-probe transient absorption spectroscopy has been performed under different pump energies. The transient absorption spectra recorded under different delay times with 2.06 and 3.1~eV pump energies are shown in Fig.~\ref{fig:TAS_abs} (a) and (b), respectively. Minima in differential absorbance $\Delta$A spectra due to the ground state bleaching associated with $A$, $B$, and $C$ excitonic features are marked in the figures. The decay profile  for the $A$ and $B$ excitonic features recorded at room temperature for the pump energies of 3.1 (Blue) and 2.06~eV (Red) are compared  in  Fig.~\ref{fig:trionformtas1}. It is important to notice that the decay times in both the cases are several hundreds of picoseconds, which  are much slower than the excitonic decay times reported in the literature \cite{robert2016exciton,faiha_2024recombination,korn2011low}. It has been found that our as-grown 1L-\mos films on sapphire substrates, where trion population is very less, show much faster decay [see supplementary material Fig.~S4 (a) \cite{sup}]. The long decay time observed in these transferred 1L-\mos films, where trion population is much higher than that of the as-grown films, can be attributed to the  trions \cite{chang_trionTRPL,gao2016valley_trionhighpower,wang_trionTRPL,yan2017long}.  It is interesting to note that when carriers are excited deep within the conduction and the valence bands with 3.1~eV of pump energy, $\Delta$A shows an initial increase for more than 20 picoseconds (indicated by dashed arrows) before decaying.  Whereas, under sub band-gap pump energy  of 2.06~eV (slightly more than the transition energy of $A$-excitons in 1L-\mos), the decay profile does not show any initial rise. Such an increase in $\Delta$A indicates the building up of the trion population in the K/K$^{\prime}$ valleys after the 3.1~eV of pump pulse. Note that the initial increasing trend is not observed in the $\Delta$A profiles in case of the as-grown samples [see supplementary material Fig.~S4 (a)]. This might be due to the presence of higher density of non-radiative recombination pathways, which restricts the number of photo-excited carriers relaxing down to K/K$^\prime$ valleys. The initial increasing effect could be observed in as-grown 1L-\mos films, when the film is vacuum annealed and capped with NiO thin film suggesting that the vacuum annealing followed by capping must be reducing the concentration of non-radiative recombination centres in the  material\cite{faiha_2024recombination}.  At low temperatures, the effect of trion build-up in the $\Delta$A decay profiles becomes even more prominent (see supplementary material Fig.~S4 (b) \cite{sup}).  In order to theoretically validate our argument, the rate equations [Eq.~(\ref{eqnn})] are numerically solved as discussed in supplementary material S5 \cite{sup}. The calculated population of excess carriers, excitons and trions  as functions of time are presented for a set of meaningful values of the coefficients in the supplementary, which indeed displays the trion build-up effect [see supplementary material Fig.~S6 \cite{sup}].

\section{Conclusions}
 In conclusion,  trions can be created in K/K$^\prime$ valleys of  1L-\mos through two different processes. When the excitation photon energy is sufficiently low, trions can be created directly in the valleys through photon absorption, which may be followed by phonon scattering events. When the excitation energy is sufficiently more than the band-gap, the carriers are generated well inside the conduction and valence bands (away from the K/K$^\prime$ valleys). These photo-excited high energy carriers release their energy via phonon emission to relax to the bottom of the K/K$^\prime$ valleys. The difference between the rate of arrival of the electrons and the holes at these valleys facilitates the transformation of the excitons into trions there. This manifests as a temporary increase of the trion population in the valley, which can be detected by the TAS study. The intensity dependent PL study under above-band-gap excitation, also carries the signature of the coexistence of the two trion generating pathways in this material. The conclusion is further validated by a rate equation model. 

\hfill{}

%\section*{Supplementary Material}
%See supplementary materials for  excitation power dependent PL spectra of 1L-\mos under different excitation energies  at low temperatures, TA spectra of the as-grown and transferred samples under different pump energies,  time variation of the transient absorption at $A$ and $B$-excitons after the pump and solutions of the rate equations.

\hfill

\section*{Data Availability}
The data that support the findings of this study are available from the corresponding author upon reasonable request.

\hfill

\section*{Acknowledgments}
We acknowledge financial support from the Science and Engineering Research Board (SERB) of the Government of India (Grant No. CRG/2022/00l852). 

\hfill

\newpage

	\Large Supplementary Materials for\\
	
	\vspace{0.2cm}
	
	\huge\textbf{Experimental demonstration of two distinct pathways of trion generation in monolayer MoS$_2$} \\
	
	\vspace{0.1cm}
	
\Large Faiha\ Mujeeb$^1$, Arkaprava\ Chowdhury$^2$, Anindya\ Datta$^2$, Subhabrata\ Dhar $^{1^*}$\\
\vspace{0.1 cm}
 $^1$Department of Physics, Indian Institute of Technology Bombay, Mumbai-400076, India\\		
$^2$Department of Chemistry, Indian Institute of Technology Bombay, Mumbai-400076, India

%		{\normalsize $^3$Department of Chemistry, Indian Institute of Technology Bombay, Mumbai-400076, India}\\
$^*$E-mail:~dhar@phy.iitb.ac.in
%	\date{\today}
	
\clearpage

%	\section*{S1. Experimental Techniques }	
%	Power dependent PL spectra were recorded at 80~K under three different excitation energies; 1.95~eV (633~nm, He-Ne Laser, ThorLabs), 2.04~eV (607~nm, DPSS Laser, Laser Glow), and 2.33~eV (532~nm, DPSS Laser, CNI Optoelectronics Technology). The data were collected in a microscopy set-up equipped with a 50$\times$ objective lens in a back scattering geometry. The spectra were recorded using a 0.55~m focal length monochromator (iHR 550, Horiba) equipped with a Peltier cooled CCD detector (Synergy Plus, Horiba).

%	Transient absorption spectroscopy (TAS) measurements were performed using a femtosecond Ti-Sapphire laser of wavelength of 800 nm (Vitesse, Coherent, USA), pulse duration of 100 fs, and repetition rate of 1 kHz. Part of the 800 nm light was passed through a CaF$_{2}$ crystal to generate a white light continuum (370–750 nm). This white light was split into probe and reference beams using a beam splitter. The reference beam was used to correct for intensity fluctuations of the probe beam. The 600~nm and 400~nm pump pulses were used for the study. The 600~nm pump was generated using an Optical Parametric Amplifier (TOPAS, Light Conversion, Lithuania), while  400~nm pump is generated via frequency multiplication of 800~nm laser. These measurements were conducted at 300~K and 80~K inside a liquid nitrogen cryostat. 

\clearpage
\renewcommand{\thefigure}{S1}
\begin{figure}[htb]
	\centering
	\includegraphics[scale=1.0]{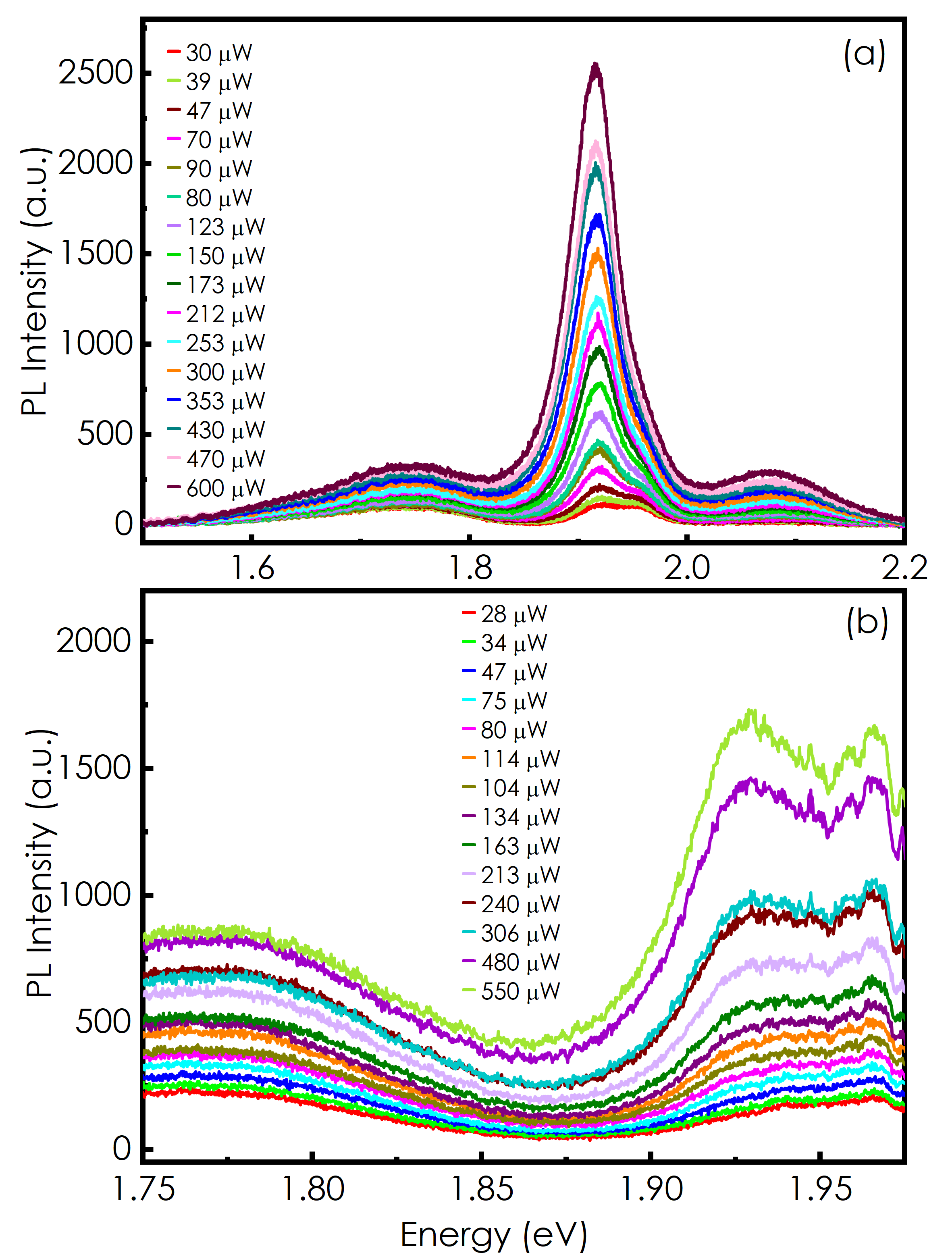}
	\caption{PL spectra recorded at 80 K under different excitation powers for excitation energies of (a) 2.33 eV (532 nm) and (b) 2.04 eV (607 nm). }
	\label{fig:PL}
	\vspace*{-\baselineskip}
\end{figure}
\newpage
\renewcommand{\thefigure}{S2}
\begin{figure}[htb]
	\centering
	\includegraphics[scale=1.0]{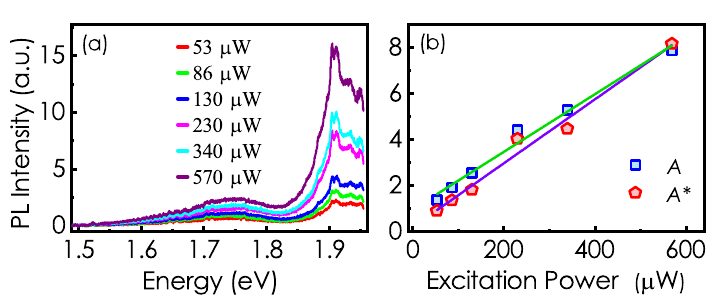}
	\caption{(a) PL spectra recorded at 100~K under different excitation powers  for excitation energy of 1.95~eV (633~nm). (b) Integrated PL intensity as a function of excitation power for excitons ($A$, blue symbols) and trions ($A^\ast$, red symbols). }
	\label{fig:PL633}
	\vspace*{-\baselineskip}
\end{figure}

\newpage
\section*{S3. Solutions of the rate equations under steady state condition}

As discussed in the main text, the steady state population of exciton ($N_{x}$), trion ($N_{t}$), and free electron ($n$) are governed by the following set of equation.

\begin{align}
	\alpha_{x}G -\gamma_{x}N_{x}+\beta_{tx}N_{t}-\beta_{xt}^{\prime}nN_{x} &=0 \label{steadystate1}\\
	\alpha_{t}nG -\gamma_{t}N_{t}-\beta_{tx}N_{t}+\beta_{xt}^{\prime}nN_{x} &=0 \label{steadystate2}\\
	\alpha_{n}G+\beta_{tx}N_{t}-\beta_{xt}^{\prime}nN_{x}-	\alpha_{t}nG &=0 \label{steadystate3}
\end{align}
Where,  $\alpha_{x}$,  $\alpha_{t}$ and  $\alpha_{n}$ are the proportionality constants associated with the generation of excitons, trions and excess carriers (electrons or holes), respectively. $\gamma_{x}$ and $\gamma_{t}$  are the recombination rate-constants for the excitons and trions, respectively.  Note that each one represents the sum of the rate-constants associated with both the radiative and nonradiative recombination processes. $\beta_{tx}$ and $\beta_{xt}^{\prime}$ are the coefficients for the rates of trion-to-exciton and exciton-to-trion conversions, respectively.  In trion-to-exciton conversion process, one trion is dissociated to form an exciton and an excess carrier. It should be noted that $n(t)$ = $n_o$ + $n_{ph}(t)$, where $n_o$ represents the background electron concentration arising from the background doping and $n_{ph}(t)$ accounts for the photo-induced change in electron concentration.

Now 
\begin{equation}
	n = \frac{\alpha_{n}G+\beta_{tx}N_{t}}{\beta_{xt}^{\prime}N_{x}+\alpha_{t}G}
	\label{n}
\end{equation}
The set of linear equations are solved in two different regimes,
\begin{enumerate}
	\item  $\alpha_{t}G > \beta_{xt}^{\prime}N_{x}$, direct trion formation dominates over exciton-to-trion conversion.
	\item $\beta_{xt}^{\prime}N_{x} > \alpha_{t}G$, exciton-to-trion conversion is the dominant process as compared to the direct formation of trions.
\end{enumerate} 
\subsection*{Case 1: $\alpha_{t}G > \beta_{xt}^{\prime}N_{x}$}
Eq.~(\ref{n}) can be written as,
\begin{equation}
	n=\frac{\alpha_{n}G+\beta_{tx}N_{t}}{\alpha_{t}G}
	\label{n_1}
\end{equation}
Eq.~(\ref{n_1}) is substituted into steady state eq.~(\ref{steadystate1}),
\begin{align}
	\alpha_{x}G -\gamma_{x}N_{x}+\beta_{tx}N_{t}-\beta_{xt}^{\prime}\frac{\alpha_{n}G+\beta_{tx}N_{t}}{\alpha_{t}G}N_{x} &=0 \\
	\alpha_{x}\alpha_{t}G^{2}-\gamma_{x}\alpha_{t}GN_{x}+\beta_{tx}\alpha_{t}GN_{t}-\beta_{xt}^{\prime}\alpha_{n}GN_{x}-\beta_{xt}^{\prime}\beta_{tx}N_{t}N_{x}&=0 \\
	-P_{1}GN_{x}+P_{2}GN_{t}-P_{3}N_{x}N_{t}+P_{4}G^{2}&=0
	\label{exciton_1}
\end{align}
where,
\begin{align}
	P_{1}&=\gamma_{x}\alpha_{t}+\beta_{xt}^{\prime}\alpha_{n} \\ 
	P_{2}&= \beta_{tx}\alpha_{t} \\
	P_{3}&=\beta_{xt}^{\prime}\beta_{tx} \\
	P_{4}&= \alpha_{x}\alpha_{t}
\end{align}
From eq.~(\ref{exciton_1}), the relation between $N_{x}$ and $N_{t}$ can be obtained,
\begin{equation}
	N_{x}=\frac{P_{4}G^{2}+P_{2}GN_{t}}{P_{1}G+P_{3}N_{t}}
	\label{eqnNx}
\end{equation}
Similarly, eq.~(\ref{n_1}) can be substituted into the rate eq.~(\ref{steadystate2}),
\begin{align}
	\alpha_{t}\left(\frac{\alpha_{n}G+\beta_{tx}N_{t}}{\alpha_{t}G}\right)G -\gamma_{t}N_{t}-\beta_{tx}N_{t}+\beta_{xt}^{\prime}\left(\frac{\alpha_{n}G+\beta_{tx}N_{t}}{\alpha_{t}G}\right)N_{x} &=0 \\
	\alpha_{t}G(\alpha_{n}G+\beta_{tx}N_{t}) -(\gamma_{t}+\beta_{tx}) \alpha_{t}GN_{t}+\beta_{xt}^{\prime}(\alpha_{n}G+\beta_{tx}N_{t})N_{x} &=0 \\
	\alpha_{n}\alpha_{t}G^{2}+\beta_{tx}\alpha_{t}GN_{t}-\alpha_{t}(\gamma_{t}+\beta_{tx})GN_{t}+\beta_{xt}^{\prime}\alpha_{n}GN_{x}+\beta_{xt}^{\prime}\beta_{tx}N_{t}N_{x} &=0 \\
	Q_{1}GN_{t}+Q_{2}GN_{x}+Q_{3}N_{x}N_{t}+Q_{4}G^{2} &=0
	\label{trion_1}
\end{align}
Where,
\begin{align}
	Q_{1}&=\beta_{tx}\alpha_{t}-\alpha_{t}(\gamma_{t}+\beta_{tx}) \\ 
	Q_{2}&= \beta_{xt}^{\prime}\alpha_{n} \\
	Q_{3}&=\beta_{xt}^{\prime}\beta_{tx} \\
	Q_{4}&= \alpha_{n}\alpha_{t}
\end{align}
The eq.~(\ref{eqnNx}) can be substituted into eq.~(\ref{trion_1}).
\begin{align}
	Q_{1}GN_{t}+Q_{2}G\left(\frac{P_{4}G^{2}+P_{2}GN_{t}}{P_{1}G+P_{3}N_{t}}\right)+Q_{3}\left(\frac{P_{4}G^{2}+P_{2}GN_{t}}{P_{1}G+P_{3}N_{t}}\right)N_{t}+Q_{4}G^{2} &=0 \\
	(Q_{1}P{3}+Q_{3}P_{2})N_{t}^{2}+(Q_{1}P_{1}+Q_{2}P_{2}+Q_{3}P_{4}+Q_{4}P_{3})GN_{t}+(Q_{2}P_{4}+Q_{4}P_{1})G^{2}&=0
\end{align}
Solution of this quadratic equation will give rise to,
\begin{equation}
	\boxed{N_{t} \propto G}
	\label{Nt_30}
\end{equation}
Substituting $N_{t} \propto G$ in eq.~(\ref{eqnNx}), 
\begin{equation}
	\boxed{N_{x} \propto G}
	\label{Nt_31}
\end{equation}

Note that the proportionality constant is different for eq. (\ref{Nt_30}) and (\ref{Nt_31}). 

\subsection*{Case 2: $\beta_{xt}^{\prime}N_{x} > \alpha_{t}G$}
The steady state free electron concentration (Eq.~(\ref{n})) can be expressed as,
\begin{equation}
	n=\frac{\alpha_{n}G+\beta_{tx}N_{t}}{\beta_{xt}^{\prime}N_{x}}
	\label{n_2}
\end{equation}
Eq.~(\ref{n_2}) is substituted into steady state eq.~(\ref{steadystate1}),
\begin{align}
	\alpha_{x}G -\gamma_{x}N_{x}+\beta_{tx}N_{t}-\beta_{xt}^{\prime}\left(\frac{\alpha_{n}G+\beta_{tx}N_{t}}{\beta_{xt}^{\prime}N_{x}}\right)N_{x} &=0 \\
	\alpha_{x}G -\gamma_{x}N_{x}+\beta_{tx}N_{t}-\alpha_{n}G-\beta_{tx}N_{t} &=0\\
	(\alpha_{x}-\alpha_{n})G -\gamma_{x}N_{x} &=0 \\
	\boxed{ N_{x} =  \frac{(\alpha_{x}-\alpha_{n})}{\gamma_{x}}G}
	\label{Nx_2}
\end{align}
The steady state exciton population is came out to be again proportional to $G$. 
In order to obtain steady state trion population with $G$, the eq.~(\ref{n_2}) is substituted into the steady state eq.~\ref{steadystate2},
\begin{align}
	\alpha_{t}\left(\frac{\alpha_{n}G+\beta_{tx}N_{t}}{\beta_{xt}^{\prime}N_{x}}\right)G -\gamma_{t}N_{t}-\beta_{tx}N_{t}+\beta_{xt}^{\prime} \left(\frac{\alpha_{n}G+\beta_{tx}N_{t}}{\beta_{xt}^{\prime}N_{x}}\right) N_{x} &=0 \\
	\alpha_{t}\left(\frac{\alpha_{n}G+\beta_{tx}N_{t}}{\beta_{xt}^{\prime}N_{x}}\right)G -\gamma_{t}N_{t}-\beta_{tx}N_{t}+ \alpha_{n}G+\beta_{tx}N_{t} &=0 \\
	\alpha_{t}\left(\frac{\alpha_{n}G+\beta_{tx}N_{t}}{\beta_{xt}^{\prime}N_{x}}\right)G -\gamma_{t}N_{t}+ \alpha_{n}G &=0 \\
	\alpha_{t}\alpha_{n}G^2+\alpha_{t}\beta_{tx}GN_{t} -\gamma_{t}\beta_{xt}^{\prime}N_{x}N_{t}+ \alpha_{n}\beta_{xt}^{\prime}GN_{x} &=0 
	\label{Nt_2}
\end{align}
The eq.~(\ref{Nx_2}) can be substituted into eq.~(\ref{Nt_2}) give rise to,
\begin{equation}
	N_{t}= \frac{\alpha_{t}\alpha_{n}\gamma_{x}+\alpha_{n}\beta_{xt}^{\prime}(\alpha_{x}-\alpha_{n})}{(\alpha_{x}-\alpha_{n})\beta_{xt}^{\prime}\alpha_{n}+\gamma_{x}\alpha_{t}\beta_{tx}} G\\ 
\end{equation}
\begin{equation}
	\boxed{N_{t} \propto G}
\end{equation}

The steady state population of both excitons and trions shows a linear relationship with generation rate.

\newpage
\renewcommand{\thefigure}{S4}
\begin{figure}[htb]
	\centering
	\includegraphics[scale=1.2]{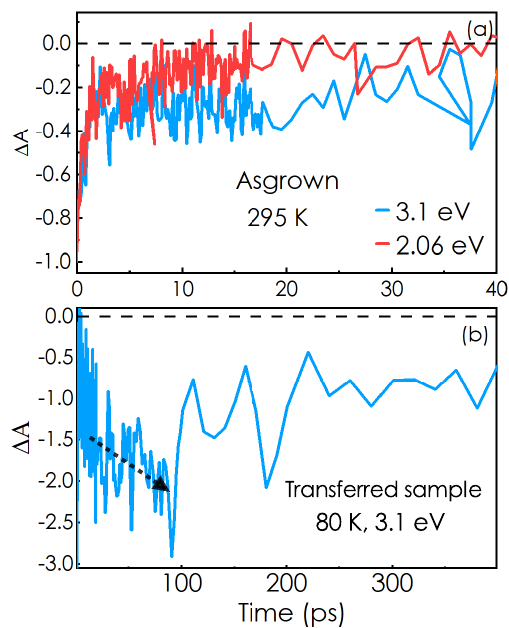}
	\caption{Normalized $\Delta$A versus delay time for (a) $A$-exciton recorded at room temperature with 3.1~eV (Blue) and 2.10~eV (Red) pump energies for the as-grown 1L-\mos on sapphire substrate and (b) $B$-excitons recorded at 80~K with 3.1~eV pump energy for the transferred sample. Dashed (black) arrows in (b) mark the increase of $\Delta$A.}
	\label{fig:Decay}
	\vspace*{-\baselineskip}
\end{figure}

\newpage
\section*{S5. Time dependent solutions of the rate equations}	

The rate equations [Eq. 1 of the main text] can be expressed as follows.  
\begin{align}
	&\frac{dN_{x}}{dt} = \alpha_{x}G -\gamma_{x}N_{x}+\beta_{tx}N_{t}-\beta_{xt}^{\prime}nN_{x}  \nonumber \\
	&\frac{dN_{t}}{dt} = \alpha_{t}nG -\gamma_{t}N_{t}-\beta_{tx}N_{t}+\beta_{xt}^{\prime}nN_{x}  \\
	&\frac{dn}{dt} = \alpha_{n}G +\beta_{tx}N_{t}-\beta_{xt}^{\prime}nN_{x}-\alpha_{t}nG \nonumber
\end{align}

$G$ = 0 in the above equations as we are interested  to know the variation of $N_{x}$,  $N_{t}$ and $n$ as functions of time after the pump pulse.  After dividing both sides of the above set of equations by $N_{0}$, the population of neutral excitons at time zero, one gets .  

\begin{align}
	&\frac{dN_{x}/N_{0}}{dt} =  -\gamma_{x}\frac{N_{x}}{N_{0}}+\beta_{tx}\frac{N_{t}}{N_{0}}-\beta_{xt}^{\prime}N_{0}\frac{n}{N_{0}}\frac{N_{x}}{N_{0}} \nonumber \\ 
	&\frac{dN_{t}/N_{0}}{dt} =  -\gamma_{t}\frac{N_{t}}{N_{0}}-\beta_{tx}\frac{N_{t}}{N_{0}}+\beta_{xt}^{\prime}N_{0}\frac{n}{N_{0}}\frac{N_{x}}{N_{0}}   \\
	&\frac{dn/N_{0}}{dt} = \beta_{tx}\frac{N_{t}}{N_{0}}-\beta_{xt}^{\prime}N_{0}\frac{n}{N_{0}}\frac{N_{x}}{N_{0}} \nonumber
\end{align} 

The above equations can be solved numerically with appropriate initial conditions to obtain the time variation of $\frac{N_{x}}{N_{0}}, \frac{N_{t}}{N_{0}}, \text{and} \frac{n}{N_{0}}$.

\newpage
\renewcommand{\thefigure}{S6}
\begin{figure}[htb]
	\centering
	\includegraphics[scale=1.0]{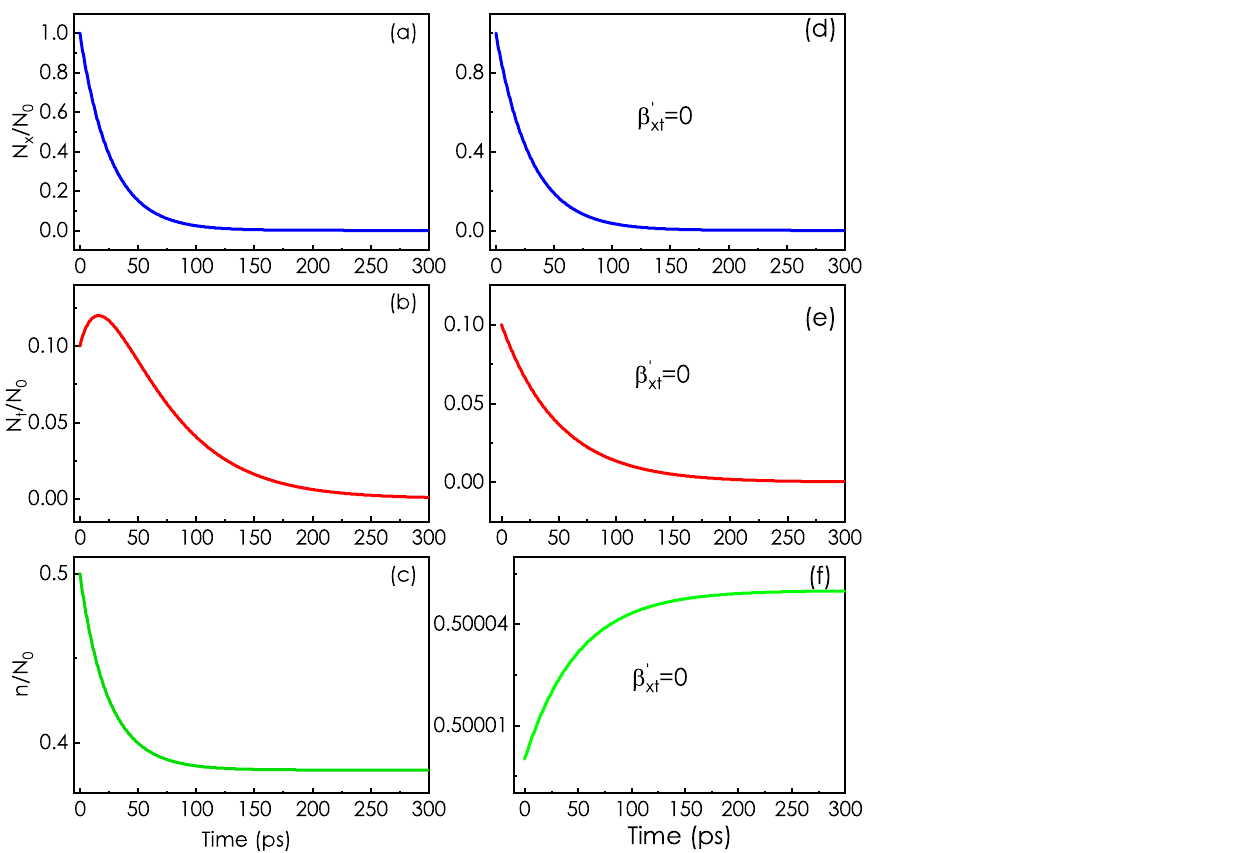}
	\caption{Time variation of $N_{x}/N_{0}$, $N_{t}/N_{0}$, and $n/N_{0}$ as obtained by solving (using Wolfram Mathematica) the rate equations for $\gamma_{x}= \frac{1}{30} \times 10^{12}$ s$^{-1}$, $\gamma_{t}= \frac{1}{100} \times 10^{12}$ s$^{-1}$, 
		%$\gamma_{n}= \frac{1}{50} \times 10^{12}$ s$^{-1}$, 
		and $ \beta_{tx} = 10^{7}$ s$^{-1}$ with initial conditions $\frac{N_{x}}{N_{0}}(t=0) = 1$, $\frac{N_{t}}{N_{0}}(t=0) = 0.1$ and  $\frac{n}{N_{0}}(t=0) = 0.5$. Panel (a,b,c) present the decay profiles when both pathway-1 and pathway-2 are active ($\beta_{xt}^{\prime}N_{0} = 10^{10}$s$^{-1}$). Panel (d,e,f) present the decay profiles  when $\beta_{xt}^{\prime} = 0$, meaning only pathway 1 is active. }
	\label{fig:solution}
	\vspace*{-\baselineskip}
\end{figure}

\bibliographystyle{unsrt}
\bibliography{bib}

\begin{thebibliography}{10}

\bibitem{mak2010atomically}
Kin~Fai Mak, Changgu Lee, James Hone, Jie Shan, and Tony~F Heinz.
\newblock Atomically thin {MoS$_2$}: a new direct-gap semiconductor.
\newblock {\em Physical review letters}, 105(13):136805, 2010.

\bibitem{mak2013tightly}
Kin~Fai Mak, Keliang He, Changgu Lee, Gwan~Hyoung Lee, James Hone, Tony~F
  Heinz, and Jie Shan.
\newblock Tightly bound trions in monolayer {MoS$_2$}.
\newblock {\em Nature materials}, 12(3):207--211, 2013.

\bibitem{mak2012control}
Kin~Fai Mak, Keliang He, Jie Shan, and Tony~F Heinz.
\newblock Control of valley polarization in monolayer {MoS$_2$} by optical
  helicity.
\newblock {\em Nature nanotechnology}, 7(8):494--498, 2012.

\bibitem{Abhay2023novel}
Abhay Anand~VS, Mihir~Kumar Sahoo, Faiha Mujeeb, Abin Varghese, Subhabrata
  Dhar, Saurabh Lodha, and Anshuman Kumar.
\newblock Novel nano-electroplating-based plasmonic platform for giant emission
  enhancement in monolayer semiconductors.
\newblock {\em ACS Applied Materials \& Interfaces}, 15(49):57783--57790, 2023.

\bibitem{Swarupdeb2022cumulative}
Swarup Deb, Wei Cao, Noam Raab, Kenji Watanabe, Takashi Taniguchi, Moshe
  Goldstein, Leeor Kronik, Michael Urbakh, Oded Hod, and Moshe Ben~Shalom.
\newblock Cumulative polarization in conductive interfacial ferroelectrics.
\newblock {\em Nature}, 612(7940):465--469, 2022.

\bibitem{qiu2013optical}
Diana~Y Qiu, Felipe~H Da~Jornada, and Steven~G Louie.
\newblock Optical spectrum of {MoS$-2$}: many-body effects and diversity of
  exciton states.
\newblock {\em Physical review letters}, 111(21):216805, 2013.

\bibitem{kheng1993observation}
K~Kheng, RT~Cox, Merle~Y d’Aubign{\'e}, Franck Bassani, K~Saminadayar, and
  S~Tatarenko.
\newblock Observation of negatively charged excitons {X}- in semiconductor
  quantum wells.
\newblock {\em Physical review letters}, 71(11):1752, 1993.

\bibitem{huard2000bound}
V~Huard, RT~Cox, K~Saminadayar, A~Arnoult, and S~Tatarenko.
\newblock Bound states in optical absorption of semiconductor quantum wells
  containing a two-dimensional electron gas.
\newblock {\em Physical review letters}, 84(1):187, 2000.

\bibitem{plechinger2015identification}
Gerd Plechinger, Philipp Nagler, Julia Kraus, Nicola Paradiso, Christoph
  Strunk, Christian Sch{\"u}ller, and Tobias Korn.
\newblock Identification of excitons, trions and biexcitons in single-layer
  {WS$_2$}.
\newblock {\em physica status solidi (RRL)--Rapid Research Letters},
  9(8):457--461, 2015.

\bibitem{berghauser2014analytical}
Gunnar Bergh{\"a}user and Ermin Malic.
\newblock Analytical approach to excitonic properties of { MoS$_2$}.
\newblock {\em Physical Review B}, 89(12):125309, 2014.

\bibitem{faiha2024solving}
Faiha Mujeeb, Vikram Mahamiya, Arushi Singh, Mansi Kothari, Arindam Chowdhury,
  Alok Shukla, and Subhabrata Dhar.
\newblock Solving the puzzle of higher photoluminescence yield at the edges of
  {MoS$_2$} monolayers grown by chemical vapor deposition.
\newblock {\em Applied Physics Letters}, 125(21), 2024.

\bibitem{jones2013optical}
Aaron~M Jones, Hongyi Yu, Nirmal~J Ghimire, Sanfeng Wu, Grant Aivazian, Jason~S
  Ross, Bo~Zhao, Jiaqiang Yan, David~G Mandrus, Di~Xiao, et~al.
\newblock Optical generation of excitonic valley coherence in monolayer
  {WSe$_2$}.
\newblock {\em Nature nanotechnology}, 8(9):634--638, 2013.

\bibitem{ross2013electrical}
Jason~S Ross, Sanfeng Wu, Hongyi Yu, Nirmal~J Ghimire, Aaron~M Jones, Grant
  Aivazian, Jiaqiang Yan, David~G Mandrus, Di~Xiao, Wang Yao, et~al.
\newblock Electrical control of neutral and charged excitons in a monolayer
  semiconductor.
\newblock {\em Nature communications}, 4(1):1474, 2013.

\bibitem{singh2014coherent}
Akshay Singh, Galan Moody, Sanfeng Wu, Yanwen Wu, Nirmal~J Ghimire, Jiaqiang
  Yan, David~G Mandrus, Xiaodong Xu, and Xiaoqin Li.
\newblock Coherent electronic coupling in atomically thin {MoSe$_2$}.
\newblock {\em Physical review letters}, 112(21):216804, 2014.

\bibitem{zhang2015valence}
Youwei Zhang, Hui Li, Haomin Wang, Ran Liu, Shi-Li Zhang, and Zhi-Jun Qiu.
\newblock On valence-band splitting in layered {MoS$_2$}.
\newblock {\em ACS nano}, 9(8):8514--8519, 2015.

\bibitem{christopher2017long}
Jason~W Christopher, Bennett~B Goldberg, and Anna~K Swan.
\newblock Long tailed trions in monolayer {MoS$_2$}: Temperature dependent
  asymmetry and resulting red-shift of trion photoluminescence spectra.
\newblock {\em Scientific reports}, 7(1):14062, 2017.

\bibitem{bellus2015tightly}
Matthew~Z Bellus, Frank Ceballos, Hsin-Ying Chiu, and Hui Zhao.
\newblock Tightly bound trions in transition metal dichalcogenide
  heterostructures.
\newblock {\em ACS nano}, 9(6):6459--6464, 2015.

\bibitem{luo2020gate}
Zhuang Luo, Hao Jia, Liu Lv, Quan Wang, and Xiaohong Yan.
\newblock Gate-tunable trion binding energy in monolayer {MoS$_2$} with
  plasmonic superlattice.
\newblock {\em Nanoscale}, 12(34):17754--17761, 2020.

\bibitem{das2020gate}
Sarthak Das, Sangeeth Kallatt, Nithin Abraham, and Kausik Majumdar.
\newblock Gate-tunable trion switch for excitonic device applications.
\newblock {\em Physical Review B}, 101(8):081413, 2020.

\bibitem{mouri2013tunable}
Shinichiro Mouri, Yuhei Miyauchi, and Kazunari Matsuda.
\newblock Tunable photoluminescence of monolayer {MoS$_2$} via chemical doping.
\newblock {\em Nano letters}, 13(12):5944--5948, 2013.

\bibitem{Faiha_AIP}
Faiha Mujeeb, Poulab Chakrabarti, and Subhabrata Dhar.
\newblock Polystyrene assisted transfer of cvd grown 1l-{MoS$_2$}: Improvement
  of optical and electrical properties.
\newblock 2995(1), 2024.

\bibitem{Faiha2023_PRB}
Faiha Mujeeb, Poulab Chakrabarti, Vikram Mahamiya, Alok Shukla, and Subhabrata
  Dhar.
\newblock Influence of defects on the valley polarization properties of
  monolayer {MoS$_2$} grown by chemical vapor deposition.
\newblock {\em Physical Review B}, 107(11):115429, 2023.

\bibitem{strain_polarization_zhu}
CR~Zhu, Gang Wang, BL~Liu, Xavier Marie, XF~Qiao, X~Zhang, XX~Wu, H~Fan,
  PH~Tan, Thierry Amand, et~al.
\newblock Strain tuning of optical emission energy and polarization in
  monolayer and bilayer {MoS$_2$}.
\newblock {\em Physical Review B}, 88(12):121301, 2013.

\bibitem{TrionPol_singh}
Akshay Singh, Kha Tran, Mirco Kolarczik, Joe Seifert, Yiping Wang, Kai Hao,
  Dennis Pleskot, Nathaniel~M Gabor, Sophia Helmrich, Nina Owschimikow, et~al.
\newblock Long-lived valley polarization of intravalley trions in monolayer
  {WSe$_2$}.
\newblock {\em Physical review letters}, 117(25):257402, 2016.

\bibitem{chang_trionTRPL}
Yia-Chung Chang, Shiue-Yuan Shiau, and Monique Combescot.
\newblock Crossover from trion-hole complex to exciton-polaron in n-doped
  two-dimensional semiconductor quantum wells.
\newblock {\em Physical Review B}, 98(23):235203, 2018.

\bibitem{gao2016valley_trionhighpower}
Feng Gao, Yongji Gong, Michael Titze, Raybel Almeida, Pulickel~M Ajayan, and
  Hebin Li.
\newblock Valley trion dynamics in monolayer {MoSe$_2$}.
\newblock {\em Physical Review B}, 94(24):245413, 2016.

\bibitem{wang2013valley_ACS}
Qinsheng Wang, Shaofeng Ge, Xiao Li, Jun Qiu, Yanxin Ji, Ji~Feng, and Dong Sun.
\newblock Valley carrier dynamics in monolayer molybdenum disulfide from
  helicity-resolved ultrafast pump-probe spectroscopy.
\newblock {\em ACS nano}, 7(12):11087--11093, 2013.

\bibitem{yan2017long}
Tengfei Yan, Siyuan Yang, Dian Li, and Xiaodong Cui.
\newblock Long valley relaxation time of free carriers in monolayer {WSe$_2$}.
\newblock {\em Physical Review B}, 95(24):241406, 2017.

\bibitem{golovynskyi2021trion}
Sergii Golovynskyi, Oleksandr~I Datsenko, Dan Dong, Yan Lin, Iqra Irfan, Baikui
  Li, Danying Lin, and Junle Qu.
\newblock Trion binding energy variation on photoluminescence excitation energy
  and power during direct to indirect bandgap crossover in monolayer and
  few-layer {MoS$_2$}.
\newblock {\em The Journal of Physical Chemistry C}, 125(32):17806--17819,
  2021.

\bibitem{lui2014trion}
CH~Lui, AJ~Frenzel, DV~Pilon, Y-H Lee, X~Ling, GM~Akselrod, J~Kong, and
  N~Gedik.
\newblock Trion-induced negative photoconductivity in monolayer {MoS$_2$}.
\newblock {\em Physical review letters}, 113(16):166801, 2014.

\bibitem{growth}
P.~K. Mohapatra, S.~Deb, B.~P. Singh, P.~Vasa, and S.~Dhar.
\newblock Strictly monolayer large continuous {MoS$_2$} films on diverse
  substrates and their luminescence properties.
\newblock {\em Appl. Phys. Lett.}, 108:042101, 2016.

\bibitem{transfer_gurarslan}
Alper Gurarslan, Yifei Yu, Liqin Su, Yiling Yu, Francisco Suarez, Shanshan Yao,
  Yong Zhu, Mehmet Ozturk, Yong Zhang, and Linyou Cao.
\newblock Surface-energy-assisted perfect transfer of centimeter-scale
  monolayer and few-layer {MoS$_2$} films onto arbitrary substrates.
\newblock {\em ACS nano}, 8(11):11522--11528, 2014.

\bibitem{chakrabarti2022enhancement}
Poulab Chakrabarti, Faiha Mujeeb, and Subhabrata Dhar.
\newblock Enhancement of valley polarization in cvd grown monolayer {MoS$_2$}
  films.
\newblock {\em Applied Physics Letters}, 121(7):072103, 2022.

\bibitem{sup}
See supplementary materials for excitation power dependent PL spectra of
  1L-{MoS$_2$} under different excitation energies at low temperatures, TA
  spectra of the as-grown and transferred samples under different pump
  energies, time variation of the transient absorption at $A$ and $B$-excitons
  after the pump and solutions of the rate equations.

\bibitem{tailoring}
S.~Deb, P.~Chakrabarti, P.~K. Mohapatra, B.~K. Barick, and S.~Dhar.
\newblock Tailoring of defect luminescence in {CVD} grown monolayer {MoS$_2$}
  film.
\newblock {\em Applied Surface Science}, 445:542--547, 2018.

\bibitem{faiha_2024recombination}
Faiha Mujeeb, Gourab Rana, Poulab Chakrabarti, Bhabani~Prasad Sahu, Rupa Jeena,
  Anindya Datta, and Subhabrata Dhar.
\newblock Recombination dynamics and manybody effect of excitons in large-area
  monolayer {MoS$_2$} capped with (111) {NiO} epitaxial layer.
\newblock {\em Journal of Physics: Condensed Matter}, 36(31):315003, 2024.

\bibitem{trion_dynamics}
Akshay Singh, Galan Moody, Kha Tran, Marie~E Scott, Vincent Overbeck, Gunnar
  Bergh{\"a}user, John Schaibley, Edward~J Seifert, Dennis Pleskot, Nathaniel~M
  Gabor, et~al.
\newblock Trion formation dynamics in monolayer transition metal
  dichalcogenides.
\newblock {\em Physical Review B}, 93(4):041401, 2016.

\bibitem{piermarocchi1997exciton}
C~Piermarocchi, F~Tassone, V~Savona, A~Quattropani, and P~Schwendimann.
\newblock Exciton formation rates in gaas/al x ga 1- x as quantum wells.
\newblock {\em Physical Review B}, 55(3):1333, 1997.

\bibitem{ceballos2016exciton}
Frank Ceballos, Qiannan Cui, Matthew~Z Bellus, and Hui Zhao.
\newblock Exciton formation in monolayer transition metal dichalcogenides.
\newblock {\em Nanoscale}, 8(22):11681--11688, 2016.

\bibitem{steinleitner2017direct}
Philipp Steinleitner, Philipp Merkl, Philipp Nagler, Joshua Mornhinweg,
  Christian Schuller, Tobias Korn, Alexey Chernikov, and Rupert Huber.
\newblock Direct observation of ultrafast exciton formation in a monolayer of
  wse2.
\newblock {\em Nano Letters}, 17(3):1455--1460, 2017.

\bibitem{trovatello2020ultrafast}
Chiara Trovatello, Florian Katsch, Nicholas~J Borys, Malte Selig, Kaiyuan Yao,
  Rocio Borrego-Varillas, Francesco Scotognella, Ilka Kriegel, Aiming Yan, Alex
  Zettl, et~al.
\newblock The ultrafast onset of exciton formation in 2d semiconductors.
\newblock {\em Nature communications}, 11(1):5277, 2020.

\bibitem{nakama2024trion}
Kota Nakama, Mitsuhiro Okada, Ryo Kitaura, Hideo Kishida, and Takeshi Koyama.
\newblock Trion formation in monolayer mos 2 observed via femtosecond
  time-resolved photoluminescence measurements.
\newblock {\em Physical Review B}, 110(23):235412, 2024.

\bibitem{robert2016exciton}
C{\'e}dric Robert, David Lagarde, Fabian Cadiz, Gang Wang, Benjamin Lassagne,
  Thierry Amand, Andrea Balocchi, Pierre Renucci, Seffattin Tongay, Bernhard
  Urbaszek, et~al.
\newblock Exciton radiative lifetime in transition metal dichalcogenide
  monolayers.
\newblock {\em Physical review B}, 93(20):205423, 2016.

\bibitem{korn2011low}
Tobias Korn, Stefanie Heydrich, Michael Hirmer, Johannes Schmutzler, and
  Christian Sch{\"u}ller.
\newblock Low-temperature photocarrier dynamics in monolayer {MoS$_2$}.
\newblock {\em Applied Physics Letters}, 99(10), 2011.

\bibitem{wang_trionTRPL}
Gamg Wang, Louis Bouet, Delphine Lagarde, Ma{\"e}l Vidal, Andrea Balocchi,
  Thierry Amand, Xavier Marie, and Bernhard Urbaszek.
\newblock Valley dynamics probed through charged and neutral exciton emission
  in monolayer {WSe$_2$}.
\newblock {\em Physical Review B}, 90(7):075413, 2014.

\end{thebibliography}

\end{document}